%% file: wormhole.tex
\documentclass[reprint,prd]{revtex4-2}

\usepackage{newtxtext,newtxmath}
\usepackage[T1]{fontenc}
\usepackage{ulem}      %

\usepackage{graphicx}
\usepackage{amsmath}	%

\usepackage[dvipsnames]{xcolor}
\usepackage{tikz}
\usepackage{hyperref}

\begin{document} 

   \title{Surface temperature of an accretion disk around a wormhole Kerr-mimicker}

   \author{A. Karakonstantakis} 
    \email{karakonang@camk.edu.pl}
    
    \author{W. Kluźniak}   %
   \affiliation{Nicolaus Copernicus Astronomical Center, ul. Bartycka 18, PL 00-716, Warsaw, Poland}

   \date{\today}

  \begin{abstract}
   It has been suggested that spinning wormholes may mimic Kerr black holes in astronomical sources such as X-ray binaries and supermassive compact objects in centers of galaxies. With recent advances in instrumentation this could be tested if clear differences between the expected wormhole and black hole accretion properties were identified.  We find that at a given circumferential radius the physical quantities relating to circular orbits in the equatorial plane are exactly the same for the spinning wormhole and a black hole of the same mass and angular momentum, if only the two space-time metrics differ in the $g_{rr}$ component alone. A Kerr-like wormhole is a perfect black hole mimicker in relation to the orbital properties in the equatorial plane. The angular velocity, specific energy, specific angular momentum, and Lense-Thirring precession rate are the same for a Kerr black hole and a Kerr-like wormhole in circular orbits of the same circumference. However, for a wormhole there are no orbits of radius less than that of its throat, and this yields an observable signature. We note that the surface area of {a disk in the equatorial plane} is different in the two spacetimes, and this allows a fairly direct method of measuring the $g_{rr}$ component of the metric when a geometrically thin {and optically thick} accretion disk is present: the radiative flux is inversely proportional to the square root of $g_{rr}$. This results in a visibly suppressed blackbody disk temperature for traversable wormholes with a sufficiently wide throat, with the temperature decreasing monotonically (relative to the Kerr result) as the throat is approached.
  \end{abstract}

    \maketitle
\section{Introduction}\label{s:intro}

Properties of several classes of astronomical objects can be understood in terms of black holes and their gravity. Nearly all galaxies have a supermassive, compact object at their centers, and direct observations with the Event Horizon Telescope \citep[][]{EHT2019, EHT2022} show that in the nearby elliptical M87, as well as in our own
Galaxy that object is no more than a few gravitational radii in size. In the latter case we also have direct observations of orbits of several stars near Sgr~A$^\ast$ that all but rule out extended supermassive bodies (such as hypothetical neutrino balls). If one excludes exotic matter, it seems clear that the compact object is a massive solution of Einstein's equations in vacuum, and a black hole is the simplest among them.

Several X-ray binaries also include an extremely compact object with no evidence for a stellar surface. The masses of these dark primaries have been measured in several cases, and at $\sim 10M_\odot$ they exceed the maximum mass of any (neutron) star modeled with an equation of state of standard nuclear matter. Their phenomenology also differs from the established neutron stars in binaries. In 
X-ray binaries the stellar-mass compact object accretes matter from a companion star, forming an accretion disk that emits intensely in the X-ray band. The thermal continuum of the disk and reflection features (such as the broad Iron K$\alpha$ line) are standard probes used to infer the 
spin of the compact object \citep{Remillard+McClintock2006,McClintock+2014}, presumed to be a black hole. 
Recently, the robustness of these spin determination methods has been tested against exotic metrics. \cite{Mummery+2024} extended the standard thin disk model to the super-extremal regime ($a > 1$), where the black hole event horizon is replaced by a naked singularity without a horizon.
Similarly, for reflection features, \cite{Mummery+Ingram2024} introduced a model to simulate iron line profiles in naked singularity spacetimes. They showed that the absence of an event horizon leads to distinct spectral signatures—such as inverted line profiles or ``triple'' lines—offering a novel test of the Cosmic Censorship Conjecture using standard X-ray observations.

Recent gravitational wave observations, provided strong evidence for the existence of black holes described by the Kerr solution. The detection of a binary black hole coalescence allows the measurement of the mass and spin of the black holes (with the observed waveform requiring no corrections to a black hole signal). In the case of GW250114 \citep{LIGO, Abac2025} the ringdown of the remnant allowed to determine its parameters as compatible with the black hole that would emerge from the merger of the constituents measured in the coalescence. 

However, in the case of X-ray binaries and the supermassive centers of galaxies we do not yet have such direct evidence. Only electromagnetic emissions are accessible at present, and it is fundamentally impossible to prove the existence of a black hole horizon from such observations alone \citep{AKL02}. A class of alternative solutions which do not have an event horizon, might mimic many properties of black holes. It is therefore important to consider the predicted motions around black hole mimickers, in the hope of excluding such alternatives observationally, taking into account motion and radiation of matter outside the putative event horizon.

One such black hole mimicker is the wormhole solution in general relativity, which contains a throat that connects two distant regions of the spacetime.
The concept of such spacetime bridges traces back to the early days of general relativity \citep{Flamm1916} and the famous Einstein-Rosen bridge \citep{Einstein+Rosen1935}. Originally regarded as non-traversable mathematical curiosities or quantum-scale fluctuations \citep{Wheeler1957}, they are now considered to be possibly existing structures. Interest in these geometries was revitalized by \cite{Morris+Thorne1988}, who defined the conditions for traversable Lorentzian wormholes. While their realization in classical general relativity requires the violation of energy conditions via ``exotic matter'' (or scalar fields, e.g., \cite{Ellis1973}), subsequent studies have extensively explored the physics of traversable wormholes (discussed in detail in 
ref.~\cite{Visser1995}) and shown that such solutions can arise naturally in various modified theories of gravity, e.g., Einstein-Gauss-Bonnet, $f(R)$, or Born-Infeld gravity, or within the context of quantum gravity (see ref. \cite{Lobo07} for a review), often without the need for additional exotic fields, e.g., \cite{Shaikh2018BI}. 
In the modern astrophysical context, these objects are increasingly studied as black hole mimickers, compact objects that lack an event horizon but may (in some cases) possess a photon sphere. They may deviate from black hole solutions only in the strong-gravity regime while remaining virtually indistinguishable in terms of matter accretion rates or electromagnetic signatures provided that a certain parameter is sufficiently small \citep{Damour+Solodukhin2007}. 

Theoretical studies suggest that if a wormhole's throat is sufficiently close to the would-be horizon radius, it can mimic the optical appearance (shadow) and orbital dynamics of a black hole \citep{Damour+Solodukhin2007}. However, subtle differences in the accretion signatures may provide a means to differentiate the wormhole from a black hole. For instance, radiation from matter traversing the throat, or modifications to the accretion disk structure due to the differing metric potential $g_{rr}$, could manifest as deviations in the X-ray spectra of X-ray binaries or 
active galactic nuclei. Distinguishing these scenarios from black hole accretion requires precise theoretical predictions of observables.

Distinguishing a Kerr black hole from such a wormhole mimicker requires identifying observables that depend on the specific deviations in the spacetime metric. Several observational channels have been proposed. 
The shadows cast by compact objects have been a primary focus, given the recently attained angular resolution of the EHT instrument \citep{EHT2022}.
Many studies investigated the effects of the wormhole throat on the shadow shape and size. 
For example, \cite{Nedkova+2013} obtained analytically the boundary of the shadow, suggesting that for low spins the wormhole shadow is nearly indistinguishable from that of a Kerr black hole. However, subsequent analyses by \cite{Shaikh2018} and specifically for the Kerr-like metric, \cite{Amir19} found that the shadow radius decreases with the deviation parameter $\lambda^2$, while \cite{Kasuya+Kobayashi2021} highlighted that for sufficient deviation parameters, the unstable photon orbit can coincide with the throat, significantly altering the shadow shape and enhancing distinguishability. Furthermore, \cite{Ohgami+Sakai2017} highlighted that the intensity contrast within the shadow region could differentiate a wormhole from a black hole. 
Images of the Kerr-like wormhole throat have been obtained in a GRMHD simulation \cite{XiaMizuno26}.
Recently it has been pointed out that for wormholes gravitational wave echoes in the post-merger ringdown signal would indicate the presence of a reflective surface or throat rather than an event horizon \citep{Cardoso17, Cardoso+Pani2017, Bueno+2018}. 
Stationary properties of accretion disks, such as their thermal emission \citep{Harko+2008, Harko+2009, Paul+2020}, and the profile of the relativistic K$\alpha$ iron line \citep{Bambi2013,LiuMizuno26}
have also been explored as ways of pinning down the nature of strong gravity.
Other methods proposed to distinguish these objects include analyzing the potentially divergent tidal forces on infalling bodies in the near-horizon limit \citep{Lemos+Zaslavskii2008}.

In this work, we focus on a specific Kerr-like rotating wormholes, originally suggested by \cite{Bueno+2018} to be an excellent Kerr mimicker when the value of a certain parameter is very small.
We find that in fact the metric perfectly mimics the equatorial orbital properties of test particles in the Kerr metric for any value of the parameter that describes the departure of the wormhole metric from the Kerr black hole one.
However, as we show, the thermal emission from a standard thin accretion disk of such wormholes exhibits unique characteristics that make it qualitatively different from the Kerr results.
By calculating the surface temperature profile of an accretion disk, we demonstrate that the modification to the radial metric component in the traversable wormhole spacetime leads to a distinct suppression of the disk temperature near the throat, providing a potentially observable signature.

\section{A Kerr-like wormhole}
\label{sec2}

We consider a stationary, axisymmetric, asymptotically flat spacetime described by the metric proposed by \cite{Bueno+2018}, 
\begin{equation}
\begin{aligned}
    ds^2 =& -(1-2Mr/\Sigma) dt^2 -2 (2 M r a \sin^2\theta / \Sigma) dtd\phi\, + \\
    &\left(r^2 + a^2 + 2 M r a^2 \sin^2{\theta}/\Sigma\right) \sin^2{\theta} d\phi^2 + \Sigma d\theta^2+\Sigma / D\, dr^2,
\end{aligned}
\label{eq:wormetric}
\end{equation}
where $M$ and $a/M$ denote the mass and dimensionless spin of the wormhole, while the functions $\Sigma,\, D$ are defined by
\begin{align*} 
    &\Sigma\equiv r^2+a^2\cos^2\theta,\\
    &D\equiv r^2-2\tilde Mr+a^2.
\end{align*}
For $\tilde M=M$
the metric reduces to the Kerr solution.
For $\tilde M\neq M$ this metric differs from the Kerr metric 
only in the radial component, $g_{rr}$. When $a\leq \tilde M$, there is a value of $r=r_\mathrm{W}$ for which $D(r_\mathrm{W})=0$. In that case a second identical copy of the space-time region $r>r_\mathrm{W}$ can be spliced at $r=r_\mathrm{W}$. For these two reasons, the spacetime described by metric (\ref{eq:wormetric}) with $a\leq \tilde M$ is referred to as a Kerr-like wormhole. The two regions $r\geq r_\mathrm{W}$ share a spherical surface at $r=r_\mathrm{W}$.
When $\tilde M>M$ we can describe the deviation of the metric from the Kerr solution with a positive parameter 
$\lambda^2$, defined through $\tilde{M} = M (1 + \lambda^2)$.
As shown in Appendix A, 
there is no event horizon when $\lambda^2>0$, the wormhole is then traversable.

It is remarkable that the orbital dynamics of massive test particles---including their angular velocity $\Omega$, specific energy $E$, and specific angular momentum $l$ in circular orbits as functions of the coordinate $r$---are identical to the Kerr case.
We present a detailed justification of this statement in Appendix~A.
However, this can already be easily seen from the form of 
the coupled equations of motion of a test particle in circular orbit in the equatorial plane, which read
\begin{equation}
    \partial_r g_{\mu\sigma}U^\mu U^\sigma=0 ,\ g_{\mu\sigma}U^\mu U^\sigma=-1,
    \label{eq:compact}
\end{equation}
and contain only the $g_{tt}$, $g_{\phi\phi}$, $g_{t\phi}$ components of the metric, because the four velocity $U$ only has nonvanishing $t$ and $\phi$ components. Thus, the orbital velocity does not depend on $g_{rr}$.

However, the radial distance, $g_{rr}^{1/2}dr$,
is different in the Kerr black hole and Kerr-like wormhole metrics. For this reason, the detailed properties of an accretion disk will differ in those two cases. In Section~\ref{s:overall} we identify an observable which in principle allows the two cases to be distinguished by astronomers.

\begin{figure*}
    \includegraphics[width=\textwidth]{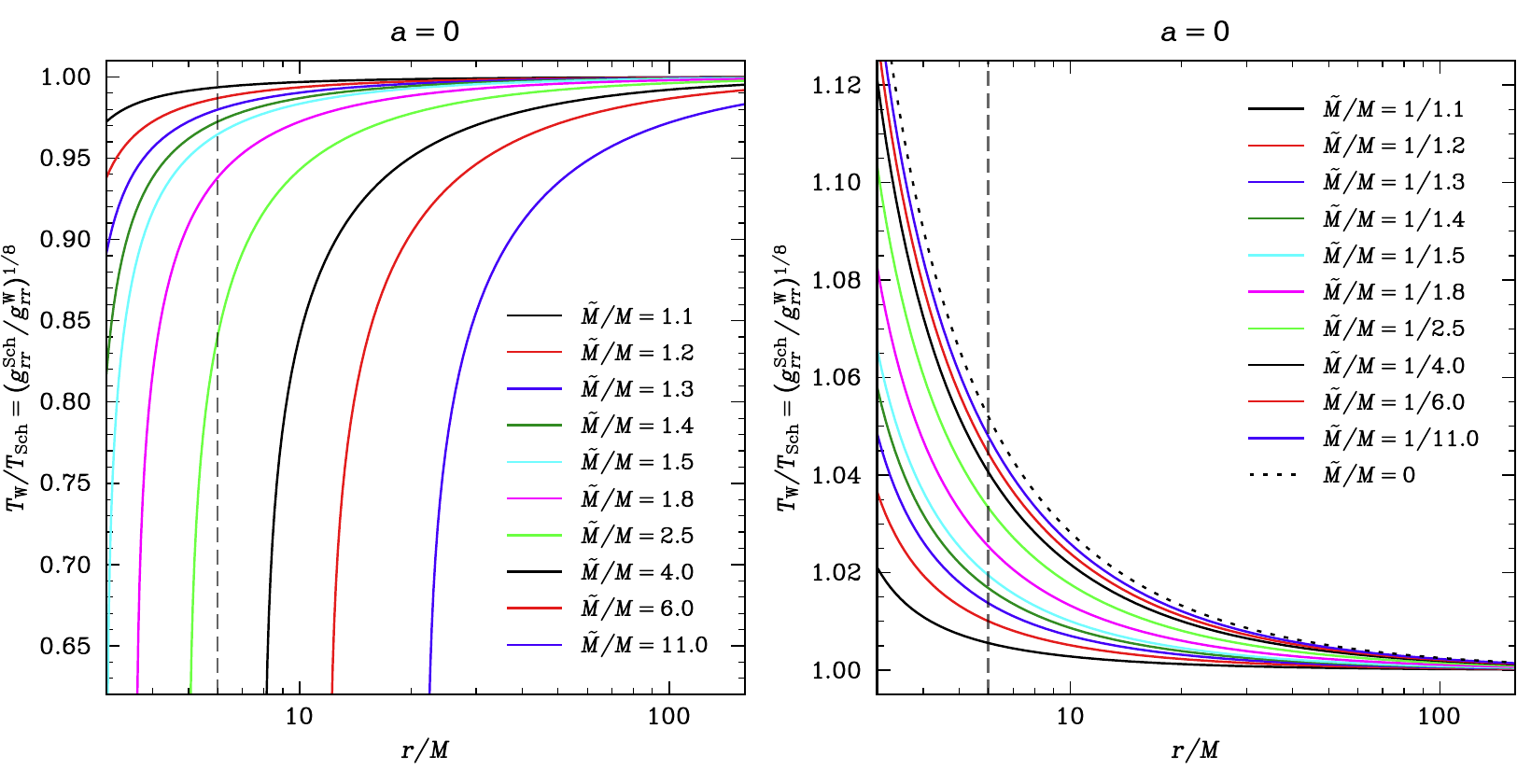}
    \caption{Ratio of the surface temperatures of an accretion disk around a non-spinning wormhole and around a Schwarzschild black hole. The vertical dashed line indicates the position of the ISCO. {\sl Left panel:} the traversable wormhole ($\tilde{M} > M$). The curves correspond to fixed values of $\tilde M/M$, which increase from left to right. {\sl Right panel:} a wormhole with an event horizon outside its throat ($\tilde{M} < M$). The curves correspond to fixed values of $\tilde M/M$, which increase from top to bottom.
    \label{schwtrav}
    \label{schwhorizon}}
\end{figure*}

\section{The overall picture}
\label{s:overall}

We begin with standard considerations of the flux emitted from the surface of an optically thick disk in the equatorial plane \citep{Shakura+Sunyaev1973}. For rotating compact objects, the Bardeen-Petterson effect \citep{Bardeen+Petterson1975} suggests that a tilted accretion disk will align its inner regions with the equatorial plane due to Lense-Thirring precession. A geometrically thin accretion disk is necessarily supported by rotation, and the fluid trajectory approximates circular orbits. Through first order, the angular velocity of the fluid is equal to that of a test particle in circular orbit \citep{Shakura+Sunyaev1973,kk2000}. The luminosity of the disk down to radius $r$ is equal to the rate at which binding  energy is released as matter drifts down from infinity to radius $r$. In the thin disk approximation the heat released by the effective viscosity is radiated locally through the disk surface, half of it escaping to the $\theta<\mathrm{\pi}/2$ hemisphere, and the other half to the $\theta>\mathrm{\pi}/2$ hemisphere.
The rate of binding energy release, down to $r$, is given by $L(r)=- [E(r)-1]\dot M$, where $E(r)$ is the specific energy of a test particle in circular orbit at radial coordinate $r$, and $\dot{M}$ is the accretion rate. We neglect here a factor of 3/2 resulting from viscous transport of angular momentum from the inner disk outwards, as it does not affect the result.
The flux from a ring between $r$ and $r+dr$ is then $F(r)=(dL/dr)/(4\mathrm{\pi}rg^{1/2}_{rr})$, the denominator being the area (per $dr$) of the two surfaces of the infinitesimal ring between $r$ and $r+dr$. 
The Newtonian limit of this formula is $F(r)=GM\dot M/(8\mathrm{\pi}r^3)$.

For an optically thick disk, blackbody radiation is a good approximation (a gray body would yield the same result).
Thus, we take $F=\sigma T^4$, where $\sigma$ is the Stefan-Boltzmann constant, or
\begin{equation}
    4\mathrm{\pi}r g^{1/2}_{rr}\sigma T^4=|dL/dr|=\dot M dE/dr.
    \label{eq:temperature}
\end{equation}
We now combine this standard expression with our result that the orbital properties of the wormhole and the Kerr black hole are identical.
The point is that the expression on the right hand side of the equation is the same for both the Kerr metric and the Kerr-like metric of Eq.~\ref{eq:wormetric}. Thus, the disk temperatures in the two metrics $g$ and $\tilde g$ satisfy the ratio
\begin{equation}
    \frac{T}{\tilde T} = \left(\frac{\tilde g_{rr}}{g_{rr}}\right)^{1/8}.
    \label{eq:Tratio}
\end{equation}
\begin{figure*}
    \includegraphics[width=\textwidth]{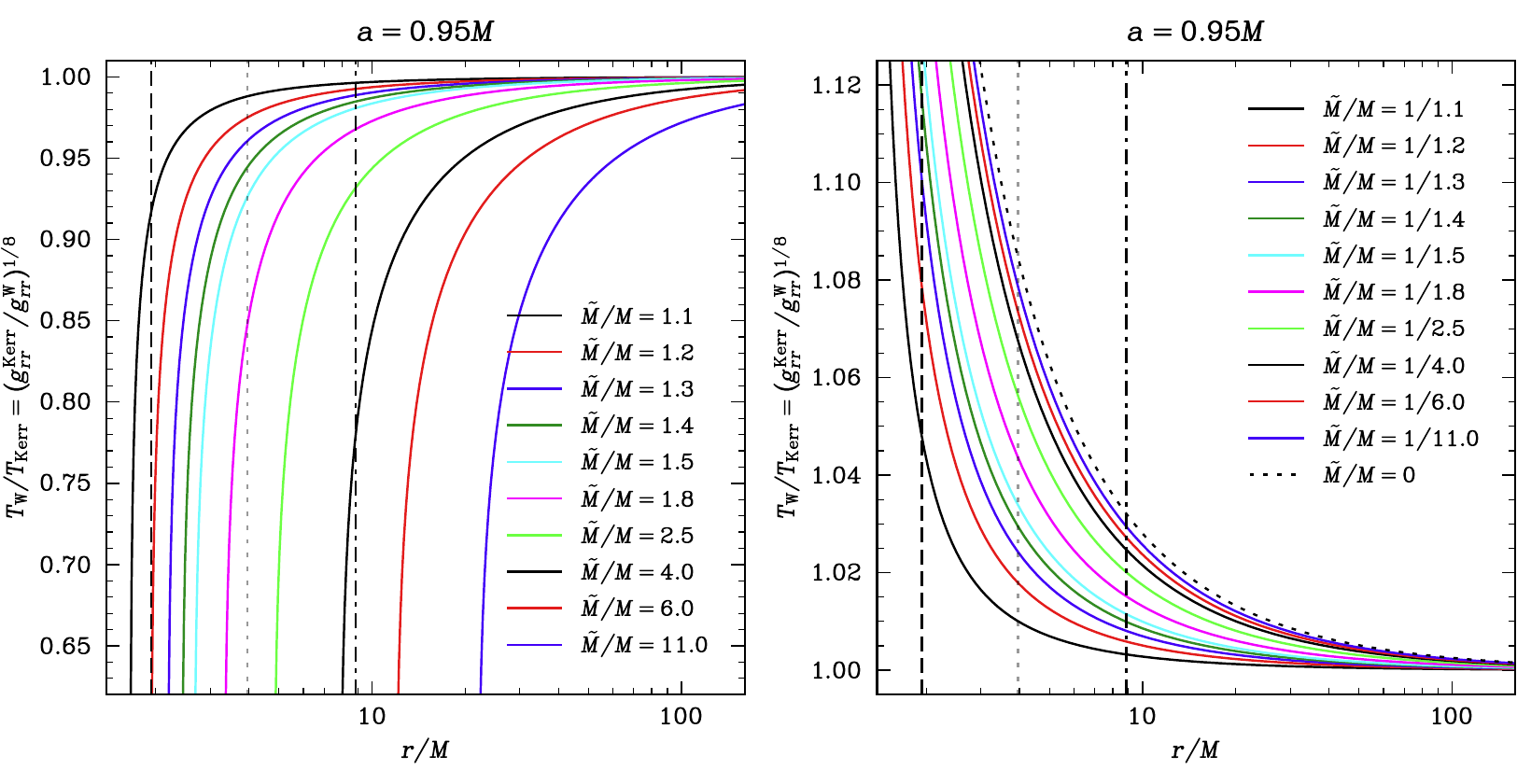}
    \caption{Ratio of the surface temperature of an accretion disk for a spinning wormhole to that of a Kerr black hole, for spin $a=0.95M$. The ordinate axis is at the photon orbit $r_\textrm{ph} = 1.386 M$. The dashed vertical line marks the radial coordinate of the ISCO. The retrograde photon orbit is at the gray dotted line, and the retrograde ISCO at the dot-dashed line. {\sl Left panel:} the traversable wormhole ($\tilde{M} > M$).  The curves correspond to fixed values of $\tilde M/M$, which increase from left to right. {\sl Right panel:} a wormhole with an event horizon outside its throat ($\tilde{M} < M$). The curves correspond to fixed values of $\tilde M/M$, which increase from top to bottom.
    \label{kerrtrav}
    \label{kerrhorizon}}
\end{figure*}
The same result follows from the standard expression \cite{PageThorne74,Balbus17} for the disk flux:
\begin{equation}
    F(r)=-\frac{\Omega'(E-\Omega l)^{-2}}{\sqrt{-\det g}}\int^r(E-\Omega l)l' dr,
\end{equation}
where $\Omega\equiv U^\phi/U^t$ is the orbital angular frequency, the specific angular momentum $l$ is defined in Appendix A, and the prime indicates a derivative over $r$. The only term depending on $g_{rr}$ is the determinant of $g$, and there are no nondiagonal $r$ components: $g_{r\nu}=0$ for $\nu\neq r$. Thus, indeed, the ratio of the fluxes at any given $r$ between the two metrics is given by 
\begin{equation}
    \frac{F}{\tilde{F}}= \frac{\sqrt{-\det \tilde{g}}}{\sqrt{-\det {g}}}=\left(\frac{\tilde{g}_{rr}}{g_{rr}}\right)^{1/2},
\end{equation}
and we assume thermal emission from an optically thick disk, $F(r)\propto [T(r)]^4$ and $\tilde{F}(r)\propto [\tilde{T}(r)]^4$, from which Eq.~\ref{eq:Tratio} follows.
This relation implies that modifications to the radial metric component directly imprint on the temperature profile. 
If the tilde denotes the Kerr black hole quantities, it is apparent that the disk temperature goes to zero at the traversable wormhole throat, where $g_{rr}$ diverges and the Kerr coefficient $\tilde g_{rr}$ is finite.

The temperature at the source determines the emitted specific intensity $I_\mathrm{disk}(T)$. In view of $I/p^3_t$ being a relativistic invariant, where $-p_t$ is the photon energy, the observed intensity is given by $\hat g^3I_\mathrm{disk}(T)$, where the $\hat g$ factor reflects the combined gravitational and Doppler shift, and is also formally independent of the radial component of the metric: $\hat g(r)=(U^\mu p_\mu)_\mathrm{obs}/(U^\mu p_\mu)_\mathrm{disk}=p_t/(p_tU^t+p_\phi U^\phi)_\mathrm{disk}$. In the stationary metrics considered here $p_t$ is conserved, so for 
any photon emitted from the disk with zero radial component of momentum, e.g. emitted
face-on 
(inclination angle $i=0$), the $\hat g$-factor 
is identical for the Kerr and Kerr-like wormholes. 
This is because $p_\phi$ is uniquely defined from the constraint $p_\mu p^\mu=0$, an equation that only involves 
metric coefficients other than $g_{rr}$, 
when $(p_r)_\mathrm{disk}=0$.
For other inclinations one needs to do ray-tracing of the photons to determine the $p_\phi$ component, and this will affect the observed spectrum.
Ray-tracing routines for the specific Kerr-like wormhole metric of Eq.~(\ref{eq:wormetric}) have been developed recently \cite{LiuMizuno26}.

\section{Results}
\label{secresults}

In Fig.~\ref{schwtrav} we compare the disk temperature of the non-spinning wormhole model ($T_\textrm{W}$) to that of the standard Schwarzschild black hole solution ($T_\textrm{Sch}$), making use of Eq.~\ref{eq:Tratio}.
The disk around the traversable wormhole appears cooler than its black hole counterpart. Each curve shows the temperature ratio of the disk of a non-spinning traversable wormhole (left panel of the figure) and a Schwarzschild black hole, i.e. when $a=0$ and $1+\lambda^2\equiv\tilde M/M>1$, for various values of the $\lambda$ parameter.

We only consider the thin disk model. Advection dominated solutions (which are more appropriate for Sgr~A$^\ast$) need to be examined more carefully---because of the greater proper radial extent of the disk the degree of advection may be altered.

While a thin disk is expected to terminate at the marginally stable orbit (the ISCO), where matter plunges towards the event horizon, or towards the throat in the case of a traversable wormhole, realistic simulations of black hole accretion at nearly Eddington rates show that the disk extends well within the ISCO, although it resembles the thin disk in its luminosity \citep[][]{puffy19,puffy22}. For this reason we terminate the plot at the photon orbit ($r=3M$ for $a=0$), rather than at the ISCO, which is denoted by the vertical dashed line in the figures. However, the plots and Eq.~\ref{eq:Tratio} do not apply to the plunge-in region, where the temperature is set by the cooling rate rather than Eq.~\ref{eq:temperature}. Whether or not the disk terminates at the ISCO, any emission from the accreting fluid on the observer's side of the wormhole portal must terminate at the throat radius, $r=r_{\mathrm{W}}$; this explains the shape of our temperature ratio curves.
The temperature suppression
persists at larger radii, and exceeds $10\%$ at the ISCO already for $\tilde M/M\approx1.5$, or $\lambda^2\approx0.5$. For somewhat larger values of $\lambda^2$ there is no marginally stable orbit and there is a 100\% suppression of the temperature at the inner edge of the disk (which coincides with the throat's equator).

For completeness, we consider also non-traversable wormholes, with an event horizon (at $r_\mathrm{H}>r_\mathrm{W}$). The temperature ratio of the disk in a wormhole and a black hole differs from unity by at most several percent and cannot exceed the value for $\tilde M=0$, as can be seen in the right panel of Fig.~\ref{schwhorizon} presenting the non-spinning case.

The results are similar for rapidly spinning objects (Fig.~\ref{kerrtrav}), the only difference being that in the case of a wormhole with a horizon the ISCO is at such low values of $r$ that the temperature enhancement may exceed 10\% for a non-traversable wormhole, and a 10\% temperature suppression in (the inner parts of) the region of stable circular motion for a traversable wormhole requires a more modest value of the wormhole parameter: $\lambda^2\approx 0.1$.

\section{Discussion and conclusions}
\label{sec:discus}
We have considered an axisymmetric, stationary, asymptotically flat space-time corresponding to a spinning wormhole of mass $M$ and angular momentum $aM$. The spacetime is Kerr-like, with a metric \citep{Bueno+2018} that differs from Kerr only in the radial, $g_{rr}$, component. Specifically, the denominator of $g_{rr}$ is $D=r^2-2\tilde Mr + a^2$, reducing to the Kerr-metric denominator $\Delta$ for $\tilde M=M$ (in which case there is no wormhole). If $\tilde M\neq M$ and $D$ has a zero (at $r=r_\mathrm{W}$), the metric describes a wormhole, which has a spherical ``throat'' at  $r=r_\mathrm{W}$ (at all times $t$). For a given value of $\tilde M$ the smallest radius of the throat is attained for the extremally spinning wormhole, $r_\mathrm{W}=\tilde M$ for $a=\tilde M$, the largest $r_\mathrm{W}=2\tilde M$ for the non-spinning wormhole ($a=0$).

If $\tilde M<M$, and $\Delta$ has a zero, the wormhole has an event horizon outside the throat. In this case the solution resembles a regular black hole \citep{Hayward2006}, with the spacetime accessible to an observer at infinity differing from Kerr only in that there is a smaller volume outside the horizon 
and correspondingly a smaller area in the equatorial plane within a given circular circumference.
Accordingly, the temperature of a geometrically thin and optically thick disk will be slightly higher than for a black hole of the same mass and spin, particularly at low radii (right panels of Figs.~\ref{schwhorizon}, \ref{kerrhorizon}).

When $\tilde M>M$ and $D$ has a zero, the metric describes a traversable wormhole. There is no horizon (and in fact no spacetime points with $r<r_\mathrm{W}$). In this case, even if $\tilde M$ is infinitesimally close to $M$, there is a fundamental difference from the Kerr black hole solution: the sphere $r=r_\mathrm{W}$ is not a one-way membrane from below which no photons or particles can emerge, instead, it is a portal to another copy of an asymptotically flat space-time. In this paper we do not discuss what can emerge from this portal, and when---one answer is given by \cite{combi24}.
Our goal is more modest: we discuss orbital properties of timelike (i.e., massive) test particles in this spacetime, on the same side of the portal (throat) as the observer at infinity, and ask what the expected X-ray flux would be for a standard thin disk.

Perhaps surprisingly, for any ratio of $\tilde M/M>1$ the properties of circular orbits of the traversable Kerr-like wormhole are identical to those of the Kerr black hole of same mass and spin. Specifically, for circular orbits in the equatorial plane, the orbital angular frequency, the energy of the test particle, its angular momentum, and the Lense-Thirring frequency have the same value at any circumferential radius in the wormhole and black hole space-times. 
A possible detection of the Lense-Thirring precession in a tidal disruption event \citep{Wang2025} allows then a determination of the spin of the compact object, but not of its nature, both a Kerr black hole and a Kerr-like wormhole being formally allowed by the measured frequency.

The position of the marginally stable orbit (ISCO) is also the same, if only it is outside the throat, $r_\mathrm{ms}>r_\mathrm{W}$. {For a given mass accretion rate and inner radius of the disk, the total luminosity of the thin disk is then identical in both spacetimes, and equal to the binding energy released down to the inner radius.} However, the area of the disk is larger for the wormhole, particularly close to the throat, this translates to a suppressed temperature of an optically thick disk at low radii (left panel of Figs.~\ref{schwtrav}, \ref{kerrtrav}). In particular, the temperature goes to zero at $r=r_\mathrm{W}$. This could be interpreted as a ``truncated disk'', reminiscent of that inferred in observations of ``black hole binaries''. Indeed, a widely accepted model of black hole X-ray emission envisions a truncated disk, with the inner thin disk replaced by a hot corona \citep{Esin+McClintock+Narayan1997, Done+2007}. The question of whether or not a hot X-ray emitting component can coexist with the thin disk outside the wormhole is beyond the scope of this work. However, for the radiatively efficient thin disks considered here, we assume the energy is radiated locally, leading to the temperature suppression we observe, and the apparent disk truncation.
It is clear that the truncation radius enforced by the spacetime geometry at $r_\mathrm{W}$ cannot change with time in the stationary metric discussed here. 

In conclusion, a thin accretion disk in the Kerr-like wormhole spacetime will not differ from that of the Kerr black hole solution in the properties related to its circular orbits, including the luminosity of any ring, angular frequency (of any blob/inhomogeneity) at a given radius, gravitational redshift %
and Doppler shifts, and Lense-Thirring precession.
In contrast, we have identified a leading 
local effect related directly to the presence of the traversable wormhole that differentiates the wormhole from the Kerr spacetime: the disk temperature will be suppressed close to the portal (throat). 
This suppression would be sufficiently substantial to be detectable ({say,\,}$\geq10$\%  correction at the inner edge of the disk) only if the wormhole parameter were on the order of unity $\lambda^2\sim1$, or more, for a non-spinning wormhole, and $\lambda^2\sim 0.1$, or more, for a rapidly spinning wormhole. For a traversable wormhole, the proper area of the disk is much larger than for a black hole, this is the reason for  suppression of emission from a thin disk at large redshifts. Hence, the inner disk temperature will be lower, and the red wings of any lines will be suppressed. The effect is particularly robust for wormholes with large throats, for an additional reason, as only radii with $r>r_\mathrm{W}$ exist in the spacetime the part of a Kerr black hole disk at $r<r_\mathrm{W}$ is missing in such a wormhole. A {general} caveat is that the different light trajectories may affect the observational appearance of the disk and its spectra---{the solid angle subtended by the disk at the observer, and the azimuthal component of the momentum of a photon  emitted with a non-zero radial component of momentum} can only be determined with ray-tracing \cite{LiuMizuno26}.

\paragraph*{Data availability.}
Software/code to generate and analyze data used in this article is publicly available at DOI:10.5281/zenodo.20796516 \cite{code}.

\acknowledgments{We thank the referees for their helpful suggestions, particularly for providing the streamlined derivation using Eq.~\ref{eq:compact}. Research supported in part by the Polish NCN grant No. 2019/35/O/ST9/039.}
\bibliographystyle{apsrev}
\input{refdef}

\bibliography{wormT} 

\appendix
\section{Circular orbits and radial distance}
Consider a stationary, axisymmetric, asymptotically flat space-time metric in standard $(t,r,\theta,\phi)$ coordinates.
\begin{equation}
    -d\tau^2=ds^2=g_{tt}dt^2 + 2g_{t\phi} dt d\phi + g_{\phi\phi}d\phi^2 +g_{\theta\theta}d\theta^2 +g_{rr}dr^2
\end{equation}
Here we show that the properties of circular orbits in the equatorial plane are unaffected by the functional form of $g_{rr}(r)$, i.e., by transformations of the
radial distance. 

Orbits of time-like test particles are easily derived from the Lagrangian
\begin{equation}
L=\frac{1}{2}\left(\frac{ds}{d\tau}\right)^2=-\frac{1}{2}
    \label{} 
\end{equation}
where the normalization corresponds to the $(-,+,+,+)$ signature.
We use the notation $\dot q\equiv dq/d\tau$.

Reflection symmetry in the equatorial plane ($\theta=\mathrm\pi/2$) of the considered space-time admits orbits in that plane. Indeed, $p_\theta=g_{\theta\theta}\dot\theta$, and necessarily $\partial L/\partial\theta|_{\theta=\mathrm{\pi}/2}=0$
under the stated assumptions, so in the equatorial plane $\dot p_\theta=0$ follows from the Euler-Lagrange equations,  and with a suitable initial condition $\theta(t)=\mathrm{const}=\mathrm{\pi}/2$ is a solution of the equations of motion.  This is the class of orbital solutions that we discuss here.

Since $\partial L/\partial t=0=\partial L/\partial \phi$, the specific energy and angular momentum, $E\equiv -p_t$ and $l\equiv p_\phi$, are constants of motion. Upon solving
\begin{equation}
    \begin{aligned}
        g_{t\phi}\dot t+g_{\phi\phi}\dot\phi&=l\\
        g_{tt}\dot t +g_{t\phi}\dot\phi&=-E
    \end{aligned}  
    \label{eq:constantmotion}
\end{equation}
for $\dot t$ and $\dot \phi$,
\begin{equation}
    \begin{aligned}
        \dot t&=(g_{\phi\phi}E+g_{t\phi}l)/\Delta,\\
        \dot\phi&=-(g_{t\phi}E +g_{tt}l)/\Delta,
    \end{aligned}
    \label{eq:dots}
\end{equation}
the constraint $(ds/d\tau)^2=-1$ yields the radial equation of motion
\begin{equation}
    g_{rr}\left(\frac{dr}{d\tau}\right)^2=V_{l,E}(r)
    \label{eq:radial}
\end{equation}
in the effective potential
\begin{equation}
    V_{l,E}(r)=(g_{tt}l^2 +2g_{t\phi}lE+g_{\phi\phi}E^2)/\Delta-1.
    \label{eq:pot}
\end{equation}
The quantity $\Delta=g_{t\phi}^2 -g_{tt}g_{\phi\phi}$ is the determinant of Eqs.~\ref{eq:constantmotion}. For the Kerr-like metric considered here, $\Delta(r,a)$ is given in Eq.~\ref{eq:delta}. 

The radial metric coefficient $g_{rr}$ does not enter the right hand side of Eqs.~\ref{eq:radial}, \ref{eq:pot}, and so the conditions for circular orbit \citep{Bardeen1972}, $V(r)=0$, $V'(r)=0$, as well as the condition for the marginally stable orbit, $V''(r)=0$, yield the same solutions, regardless of the functional form of $g_{rr}$. This important conclusion can be derived formally.
We note that if 
\begin{equation}
    \begin{aligned}
        V(r)=0, \\
        V'(r)=0
    \end{aligned}
    \label{condition}
\end{equation}
at some value $r=r_0$, and if $u$ is a function of $r$, then the function\footnote{This freedom of multiplying Eq.~\ref{eq:radial} by any function of $r$ makes the exact functional form of the effective potential a matter of convenience. The familiar effective potential of \cite{Bardeen1972} is given by $V_r=(\Sigma\Delta) V_{l,E}$.} $\bar U(r)=u(r)V(r)$ satisfies the same conditions at the radius $r=r_0$:
\begin{equation}
    \begin{aligned}
        \bar U(r)=0,\\
        \bar U'(r)=0.
    \end{aligned}
    \label{eq:zeroes}
\end{equation}
This is because, obviously, $\bar U'=u'V +u V'=0$ at $r=r_0$, by virtue of relations \ref{condition}. Similar considerations apply to $V''=0$ and $\bar U''=0$.

{\sl Corollary.} Two metrics that differ only in the $g_{rr}(r)$ coefficient admit the same circular orbits. 

Indeed, if the radial coefficients are $g_{rr}$ and $\tilde g_{rr}$, the effective potentials of the two metrics, and their first two derivatives have the same zeroes by Eq.~\ref{eq:zeroes} with $u\equiv \tilde g_{rr}/g_{rr}$, as Eq.~\ref{eq:radial} written in the form
\begin{equation}
    \tilde g_{rr}\left(\frac{dr}{d\tau}\right)^2 =u(r)V_{l,E}(r).
\end{equation}

For the Kerr-like wormhole considered here, this implies that, as a function of the coordinate $r$, the specific energy $E$ and angular momentum $l$ in circular orbits, the radius of the ISCO, and the angular velocity in circular orbits are the same as the Kerr-metric expressions derived in \cite{Bardeen1972}.  Specifically, the values of energy and angular momentum of a test particle in circular orbit, $E(r)$ and $l(r)$, can be determined as functions of the orbital radius from the conditions  $V(r)=0$, $V'(r)=0$ \citep{Bardeen1972}. 
For instance, the angular velocity with respect to a distant observer of a test time-like particle in circular orbit is given by 
\begin{equation}
    \Omega_{\rm K}(r) = \frac{M^{1/2}}{r^{3/2}+a M^{1/2}}.
\end{equation}
We use geometrized units ($G=1$, $c=1$), $M$ is the mass of the source, and $aM$ its angular momentum.
Another useful expression gives the specific energy (per unit rest mass) of a test particle in circular orbit \citep{Bardeen1972}:
\begin{equation}
    E (r)= \frac{1 - 2 M / r \pm aM^{1/2} / r^{3/2}}{\left(1 - 3M / r \pm 2aM^{1/2} / r^{3/2}\right)^{1/2}}\ ,
\end{equation}
where, the upper (lower) sign, $+a$ ($-a$), refers to prograde (retrograde) orbits.
The Newtonian limit,  $E=1-M/(2r)$, is obtained by neglecting terms of order $(M/r)^{3/2}$ and higher. 
 
Circular orbits can only be observed outside the event horizon, if present. If a wormhole throat is present, no spacetime point has a value of the radial coordinate lower than that of the throat. We must, therefore, make sure that any circular orbit discussed here  has radial coordinate larger than that of the wormhole throat or event horizon, if present.

For the Kerr metric in Boyer-Lindquist coordinates \citep{Bardeen1972}
\begin{equation}
    \begin{aligned}
        g_{tt}&=-(1-2Mr/\Sigma),  &  g_{t\phi} &= -2 a M r \sin^2\theta / \Sigma, \\
        g_{\theta\theta} &= \Sigma,&g_{rr} &= \Sigma / \Delta, &
        \label{eq:Kerr}
    \end{aligned}
\end{equation}
$ g_{\phi\phi} = \left(r^2 + a^2 + 2 Mr a^2  \sin^2{\theta}/\Sigma\right)\sin^2{\theta},$\\
with
\begin{equation} 
    \begin{aligned}
        &\Sigma\equiv r^2+a^2\cos^2\theta,\\
        &\Delta\equiv r^2-2Mr+a^2,\\
    \end{aligned}
    \label{eq:delta} 
\end{equation}
while for the Kerr-like wormhole we use \citep{Bueno+2018}
\begin{equation}
    \begin{aligned}
       g_{tt}&=-(1-2Mr/\Sigma), & g_{t\phi} &= -2 a M r \sin^2\theta / \Sigma, \\
        g_{\theta\theta} &= \Sigma, 
    &g_{rr} &= \Sigma / D,
        \label{eq:Kerrlike}
    \end{aligned}
\end{equation}
$g_{\phi\phi} = \left(r^2 + a^2 + 2 Mr a^2  \sin^2{\theta}/\Sigma\right) \sin^2{\theta}$\\
with the functions of Eq.~\ref{eq:delta}, and
\begin{equation}
    D\equiv r^2-2\tilde Mr+a^2.
    \label{eq:d}
\end{equation}
Thus the only difference between the Kerr and the Kerr-like metric is in the $g_{rr}$ term, and that only if $\tilde M\neq M$.

At the outer root of $D$, i.e. at $r_\mathrm{W}=\tilde M+\sqrt{\tilde M^2-a^2}$, the inverse of $g_{rr}$ vanishes for the metric of Eq.~\ref{eq:Kerrlike}, and if at this radius another copy of the space-time is attached, the metric describes a wormhole, with $r_\mathrm{W}$ the radial coordinate of the spherical throat, or portal to another asymptotically flat region.
For such a spacetime, the coordinate $r$ is restricted to values larger than or equal to the radius of the throat. 
In this paper we are only interested in one copy of the spacetime region with $r> r_\mathrm{W}$, the observer and the accretion disk are in the same asymptotically flat region.

We note that for the stationary space-time considered here, the event horizon, if it exists, corresponds to a null hypersurface of $r=\mathrm{const}$. Obviously, the value of $g_{rr}$ does not affect the position or the shape of the event horizon, as $dr=0$ for that hypersurface. The event horizon, if it exists, is given by the condition $\Delta=0$ (outer root), i.e. $r_\mathrm{H}=M+\sqrt{M^2-a^2}$ for both metrics Eqs.~\ref{eq:Kerr}, \ref{eq:Kerrlike}. 
In case of doubt, one can inspect Eqs.~\ref{eq:standard}, \ref{eq:healing}. Thus, for the spacetime described by the metric~\ref{eq:Kerrlike} the event horizon only exists if it is outside the wormhole throat,  $r_\mathrm{H}>r_\mathrm{W}$; such a wormhole is not traversable: no null or time-like particles can emerge from within the event horizon. If the outer root of $\Delta$ is smaller than the outer root of $D$, then the event horizon cannot exist, because it is outside the spacetime domain, $r_\mathrm{H} \notin(r_\mathrm{W}, \infty)$ in that case. From the expressions \ref{eq:delta} and \ref{eq:d} we see that the wormhole is traversable (it lacks an event horizon) if and only if $\tilde M>M$. Sometimes this is written as $\tilde M=M(1+\lambda^2)$.

To discuss stability of motion in the equatorial plane (against perturbations in $\theta$) it is convenient to use the standard metric form of an axisymmetric, stationary, asymptotically flat spacetime:
\begin{equation} 
ds^2=-e^{2\nu}dt^2+ e^{2\psi}(d\phi-\omega dt)^2 + e^{2\mu_2}d\theta^2 + e^{2\mu_1}dr^2.
    \label{eq:standard} 
\end{equation}
For the Kerr-like spacetime
\begin{equation}
    \begin{aligned}
       e^{2\nu}&=\Sigma\Delta/A,  & e^{2\psi} &=\left(\sin^2\theta\right)A/\Sigma, 
        \\
         e^{2\mu_2}&=\Sigma,  & e^{2\mu_1}&=\Sigma/D,
        \label{eq:healing}
    \end{aligned}
\end{equation}
with Eqs.~\ref{eq:delta}, \ref{eq:d}, and
\begin{equation} 
    \begin{aligned}
        &A\equiv (r^2+a^2)^2-a^2\Delta\sin^2\theta ,\\
        &\omega=2Mar/A.\\
    \end{aligned}
    \label{} 
\end{equation}

For the Kerr or Kerr-like metric of Eqs.~\ref{eq:Kerr}, \ref{eq:Kerrlike} orbits in the equatorial plane are stable against out-of-plane perturbations because $\partial L/\partial\theta \propto \cos\theta$ with a positive proportionality coefficient, so $\dot p_\theta$ is directed towards the plane $\theta=\mathrm{\pi}/2$:
\begin{equation} 
\frac{\partial L}{\partial\theta}=[c_0 \dot t^2+c_1 (\dot\phi-\omega\dot t)^2+c_2\dot\theta^2+c_3\dot r^2]\sin 2\theta,
    \label{eq:stability} 
\end{equation}
with the functions $c_i(r,\theta)>0$, $i=0,1,2,3$. For instance, $c_0=2 \Delta A^{-2}Mra^2(r^2+a^2)$, and $c_1=(A/\Sigma)  +\sin^2\theta(A/\Sigma)^2c_0/\Delta$.
Again, we note that for a circular orbit ($\dot r=0$) the right hand side of Eq.~\ref{eq:stability} does not involve the $g_{rr}$ component. It is already clear that by dint of the Euler-Lagrange equations small out-of-plane excursions lead to a harmonic oscillator equation in $\theta$ that does not depend on $g_{rr}$. Thus, the vertical epicyclic frequency for the Kerr-like wormhole is given by the Kerr-metric expression
\begin{equation}
    \omega_\perp(r)=\Omega_\mathrm{K}\sqrt{1-\frac{4\,a M^{1/2}}{r^{3/2}} +
    \frac{3a^2}{r^{2}}}.
\end{equation}
The Lense-Thirring precession rate, being equal to $\Omega_{\rm LT}=\Omega_\mathrm{K}
-\omega_\perp$, is also the same for a Kerr black hole and the Kerr-like wormhole.

\end{document}

%% file: refdef.tex
\def\prc{Phys. Rev. C }
\def\pre{Phys. Rev. E }
\def\prd{Phys. Rev. D }
\def\jcap{Journal of Cosmology and Astroparticle Physics }
\def\apss{Astrophysics and Space Science }
\def\mnras{Monthly Notices of the Royal Astronomical Society }
\def\apj{The Astrophysical Journal }
\def\aap{Astronomy and Astrophysics }
\def\aapr{Astronomy and Astrophysics Review }
\def\actaa{Acta Astronomica }
\def\pasj{Publications of the Astronomical Society of Japan }
\def\apjl{Astrophysical Journal Letters }
\def\pasa{Publications Astronomical Society of Australia }
\def\nat{Nature }
\def\physrep{Physics Reports }
\def\araa{Annual Review of Astronomy and Astrophysics}
\def\apjs{The Astrophysical Journal Supplement}
\def\na{New Astronomy}

\def\mdash{---}